\documentclass[conference]{IEEEtran}
\IEEEoverridecommandlockouts
\usepackage{cite}
\usepackage{amsmath,amssymb,amsfonts}
\usepackage{tabularx}
\usepackage{makecell}
\usepackage{algorithmic}
\usepackage{graphicx}
\usepackage[bookmarks=false]{hyperref}

\usepackage{textcomp}
\usepackage{xcolor}
\usepackage{booktabs}
\usepackage{multirow}
\usepackage{rotating}
\def\BibTeX{{\rm B\kern-.05em{\sc i\kern-.025em b}\kern-.08em
    T\kern-.1667em\lower.7ex\hbox{E}\kern-.125emX}}
\begin{document}

\title{AgenticPrecoding: LLM-Empowered Multi-Agent System for Precoding Optimization}

\author{

\IEEEauthorblockN{Zijiu Yang, Zixiang Zhang, Shunpu Tang, Qianqian Yang, Zhiguo Shi }
\IEEEauthorblockA{ 
College of Information Science and Electronic Engineering, Zhejiang University, Hangzhou, China \\
Email: \{zijiu\_yang, zhangzixiang, tangshunpu, qianqianyang20, shizg\}@zju.edu.cn
}
\thanks{
This work is partly supported by the National Key R\&D Program of China under Grant 2024YFE0200802, by the NSFC under grant No. 62293481 and No. 62571487,  and by the Zhejiang Provincial Natural Science Foundation of China under Grant No. LZ25F010001.  (Corresponding author: Q. Yang.)
}
}

\maketitle

\begin{abstract}
 Precoding is a key technique for interference management and performance improvement in multi-antenna wireless systems. However, existing precoding methods are typically developed for specific system models, objectives, and constraint sets, which limits their adaptability to the heterogeneous and evolving scenarios expected in future 6G networks. To address this limitation, we propose \textbf{AgenticPrecoding}, a universal multi-agent framework that automates end-to-end precoding derivation directly from user-level communication requirements. Specifically, \textbf{AgenticPrecoding} decomposes the derivation into four coordinated stages: problem formulation, solver selection, prompt upsampling, and code generation, and assigns each stage to an agent suited to its reasoning demand. We employ two LoRA-adapted reasoning agents to inject precoding-specific knowledge for problem formulation and solver selection, while two general-purpose LLMs handle prompt refinement and executable code generation. Moreover, a feedback-driven refinement mechanism is further incorporated to improve code executability, constraint feasibility, and solution quality. Experiments on 10 representative precoding scenarios demonstrate that the proposed \textbf{AgenticPrecoding} achieves superior cross-scenario adaptability over conventional optimization-based and LLM-based baselines.

\end{abstract}


\section{Introduction}

Over the past decades, multi-antenna technology has become a key enabler of modern wireless communications, bringing large gains in spectral efficiency and spatial multiplexing. As one of its core signal processing techniques, precoding shapes the transmitted signal across antennas to reduce interference and focus energy on intended users, and has been widely adopted in current 5G massive multiple-input multiple-output (MIMO) systems. However, precoding design in future sixth-generation (6G) networks is expected to be much more challenging. Specifically, 6G will cover highly diverse scenarios, such as non-terrestrial networks (NTNs), integrated sensing and communication (ISAC) with different antenna architectures and service requirements. As a result, precoding can no longer be designed for a few fixed models, but must work reliably across a wide range of channel conditions, system scalability, and various constraints, which makes \emph{general-purpose precoding design} an important yet unsolved problem.

\begin{figure}[t!]
	\begin{center}
		\centerline{\includegraphics[width=\columnwidth]{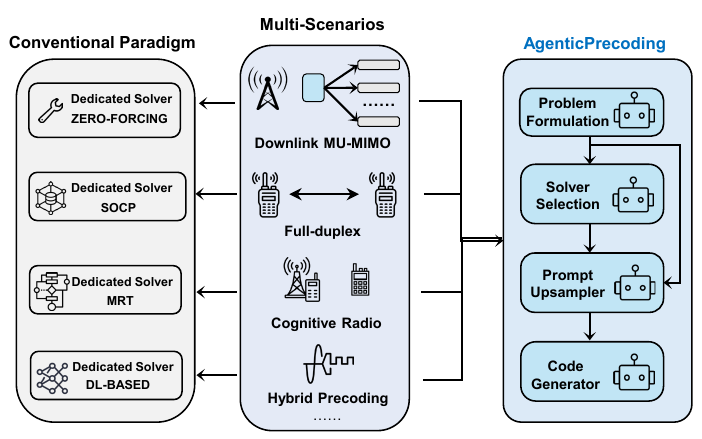}}
		\caption{Comparison between the conventional scenario-specific precoding paradigm and the proposed AgenticPrecoding framework, which transforms diverse wireless precoding scenarios into an automated multi-agent workflow including problem formulation, solver selection, prompt upsampling, and code generation.}
		\label{graph1}
	\end{center}
\end{figure}

However, existing precoding methods largely follow a ``one scenario, one formulation, one solver'' paradigm. Specifically, customized precoders have been developed for full-duplex \cite{FD_SE_EE,FD_power}, downlink MIMO \cite{DL_low,DL_power,DL_reinf,DL_robustness,DL_security,DL_sumrate}, cognitive radio \cite{CR_inf,CR_robust,CR_security,CR_sumrate}, constant-envelope \cite{CE_CI,CE_power}, distributed-antenna \cite{DA}, spatio-temporal symbol-level \cite{ST}, and hybrid precoding \cite{HP_CI,HP_robust} systems, each relying on dedicated modeling and solver derivation. Although recent deep learning (DL)-based methods can learn precoding mappings directly from large dataset \cite{DeL1,DeL2,DeL3,DeL4} and reduce the need for problem-specific derivation, they typically degrade once deployment conditions differ from the training setup and also do not explicitly capture the mathematical structure of precoding problems. As a result, both paradigms offer limited adaptability to the heterogeneous scenarios envisioned for 6G.

In recent years, large language models (LLMs) have demonstrated remarkable success across diverse tasks, enabled by scaling-law-driven improvements in model capability, which has led to growing interest in using LLMs to solve problems in the physical layer of wireless communications. Specifically, the authors in \cite{llm-optira} proposed LLM-Optira, a single CoT-based agent that directly solves optimization problems. In addition, BeamAgent \cite{beamagent} and LLM-beamforming \cite{llm_precoding_op} leverage LLMs to translate user requirements into optimization parameters or to serve as the backbone of scenario-specific beamforming design. However, these methods either suffer from limited implementation reliability and frequent feasibility violations, or remain tied to scenario-specific optimizers, and therefore leave universal precoding automation largely unresolved.



To fill this gap, we aim to build a unified framework that automatically derives reliable precoding solutions across diverse wireless scenarios, directly from user-level task descriptions. However, such a framework involves multiple reasoning stages with different needs: problem formulation and solver selection require precoding-specific knowledge, while prompt upsampling and code generation rely more on general language and coding ability. Motivated by this, we propose \textit{AgenticPrecoding}, a multi-agent framework that decomposes the end-to-end derivation into four coordinated stages, namely problem formulation, solver selection, prompt upsampling, and code generation, and matches each stage with an agent suited to its reasoning need, as illustrated in \autoref{graph1}. The main contributions of this paper are summarized as follows:
\begin{itemize}
    \item We propose \emph{AgenticPrecoding}, a universal multi-agent framework that automates end-to-end precoding derivation across heterogeneous wireless scenarios, directly from user descriptions.
    \item We design a multi-agent architecture, in which the \emph{Problem Formulator Agent} and \textit{Problem Solver Agent} are LoRA-fine-tuned to inject precoding-specific knowledge, while the \textit{Prompt Upsampler Agent} and \textit{Code Generation Agent} are instantiated with general-purpose LLMs. A feedback-driven refinement mechanism is further incorporated to improve the executability, feasibility, and solution quality of the generated solver.
    \item We conduct extensive experiments on 10 representative precoding scenarios. The results show that the proposed \emph{AgenticPrecoding} consistently achieves strong performance and cross-scenario adaptability compared with conventional optimization-based baselines.
\end{itemize}

\section{Problem Formulation}

\subsection{Generalized System Model}
In this paper, we consider a unified precoding optimization model that can encompass a broad class of representative wireless precoding scenarios. Specifically, a generic precoding optimization problem can formulated as
\begin{equation}
\label{eq:general_problem}
    \begin{aligned}
        \min_{\mathbf{x} \in \mathcal{X}(\Theta)} \quad 
        & f(\mathbf{x}; \Theta) \\
        \mathrm{s.t.} \quad 
        & c_\ell(\mathbf{x}; \Theta) \leq 0, \quad \ell = 1,\dots,L, \\
        & h_q(\mathbf{x}; \Theta) = 0, \quad q = 1,\dots,Q,
    \end{aligned}
\end{equation}
where $\mathbf{x}$ denotes the design variable, $\mathcal{X}(\Theta)$ denotes the scenario-dependent feasible domain, $f(\cdot;\Theta)$ is the objective function, and $\{c_\ell(\cdot;\Theta)\}_{\ell=1}^{L}$ and $\{h_q(\cdot;\Theta)\}_{q=1}^{Q}$ represent the inequality and equality constraints, respectively.

To be more specific, the scenario descriptor $\Theta$ collects the parameters that characterize a particular precoding setup, given by
\begin{equation}
    \Theta = \{\Theta_{\mathrm{sys}}, \Theta_{\mathrm{ch}}, \Theta_{\mathrm{obj}}, \Theta_{\mathrm{con}}\}.
\end{equation}
where $\Theta_{\mathrm{sys}}$ specifies the system configuration, such as network topology, transceiver architecture, and antenna dimensions, $\mathbf{x}$ may denote beamforming matrices, symbol-level transmit vectors, or auxiliary variables introduced by reformulation. Moreover, $\Theta_{\mathrm{ch}}$ denotes the channel-related information, such as channel matrices, interference links, and $\Theta_{\mathrm{obj}}$ defines the communication-oriented design objective, such as power minimization, sum-rate maximization, spectral-efficiency maximization, interference suppression. $\Theta_{\mathrm{con}}$ characterizes the operational constraints, including power budgets, quality-of-service requirements, interference-temperature limits, symbol-region constraints, hardware restrictions, and architecture-dependent feasibility conditions. We note that for maximization tasks, \eqref{eq:general_problem} is obtained by minimizing the negative of the corresponding performance metric. 

\subsection{Task Formulation of AgenticPrecoding}

Based on the generalized system model above, \emph{AgenticPrecoding} is formulated as an automated solver-generation task for user-specified wireless precoding scenarios. The input to the framework consists of a natural-language task description $D$ and the corresponding structured scenario descriptor $\Theta$, i.e.,
\begin{equation}
    \mathcal{I} = (D,\Theta).
\end{equation}
where $D$ describes the communication-level design requirement, such as power minimization, spectral-efficiency maximization, interference suppression, robustness enhancement, or secrecy improvement, while $\Theta$ provides the structured system, channel, objective, and constraint information needed to instantiate the corresponding optimization problem.

Given $\mathcal{I}$, the objective of \emph{AgenticPrecoding} is to infer a complete solver-generation pipeline
\begin{equation}
    \mathcal{G}: (D,\Theta) \mapsto (\mathcal{P}, \mathcal{A}, C, \mathbf{x}^{\star}),
\end{equation}
where $\mathcal{P}$ denotes the mathematical optimization problem instantiated from \eqref{eq:general_problem}, $\mathcal{A}$ denotes the selected solution algorithm, $C$ denotes the generated executable solver code, and $\mathbf{x}^{\star}$ denotes the obtained precoding solution. In other words, \emph{AgenticPrecoding} transforms high-level communication requirements and structured system parameters into a mathematically consistent optimization formulation, selects an appropriate algorithm according to the problem structure and constraints, and generates executable code to derive a feasible precoding solution without manual intervention.
\begin{figure*}[!ht]
	\centering
	\includegraphics[width=\textwidth]{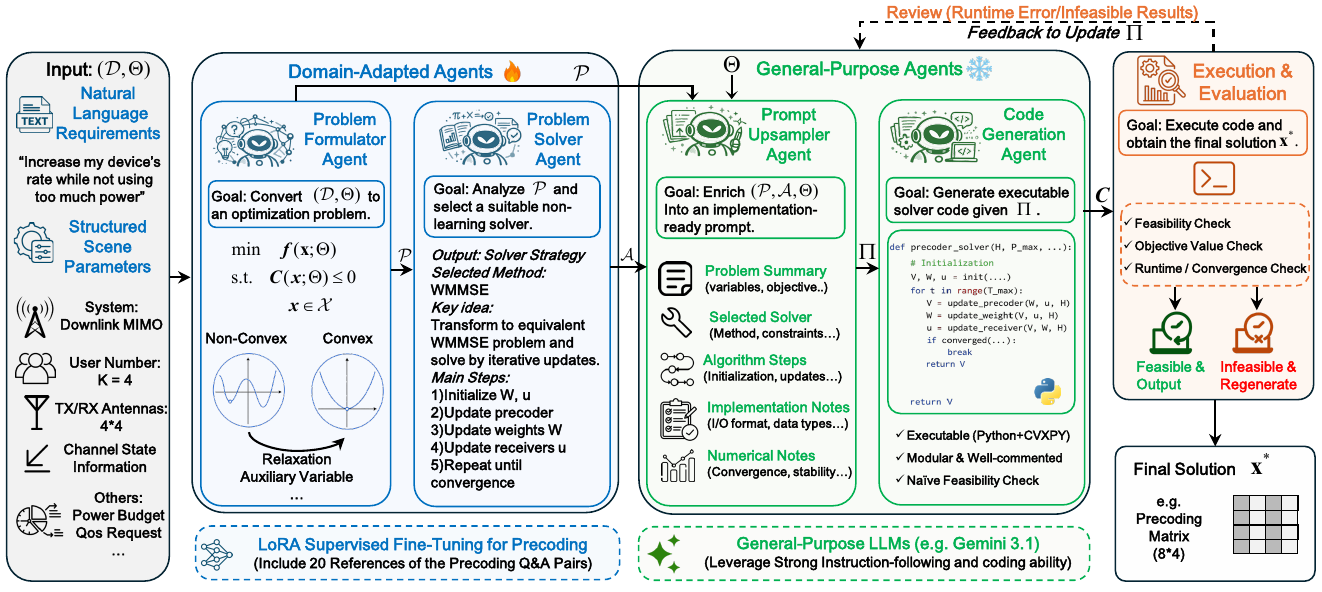}
	\caption{Illustration of the proposed AgenticPrecoding framework, with the collaborative multi-agent workflow that transforms user-level requirements and system parameters into executable solver code through four stages: problem formulation, solver selection, prompt upsampling, and code generation, integrated with a feedback-driven refinement mechanism.
}
	\label{graph2}
\end{figure*}

\section{Proposed Framework}

In this section, we present the proposed \emph{AgenticPrecoding} framework for automated precoding optimization. As illustrated in \autoref{graph2}, the framework decomposes the solver-generation process into four stages: problem formulation, solver selection, prompt upsampling, and code generation. This design avoids requiring a single LLM to simultaneously handle wireless-domain modeling, algorithmic reasoning, implementation planning, and code synthesis.

\subsection{Framework Overview}
Given the user requirement $D$ and scenario descriptor $\Theta$, \emph{AgenticPrecoding} first employs two domain-adapted reasoning agents to produce a structured problem specification and select an appropriate solver strategy. The resulting information is then passed to two general-purpose implementation agents, which reorganize the problem and solver information into an implementation-oriented prompt and generate executable solver code. Different from a one-shot generation pipeline, \emph{AgenticPrecoding} further incorporates a feedback-driven refinement loop after code execution. Specifically, the execution results are evaluated in terms of running status, constraint feasibility, and solution quality, and the resulting diagnostic feedback is used to revise the generated code when implementation errors, infeasible outputs, or unsatisfactory objective values are detected. The overall workflow can be summarized as
\begin{equation}
\begin{aligned}
    (D,\Theta)
    &\xrightarrow{\mathrm{PF}}
    \mathcal{P}
    \xrightarrow{\mathrm{PS}}
    \mathcal{A}, \\
    (\Theta,\mathcal{P},\mathcal{A})
    &\xrightarrow{\mathrm{PU}}
    \Pi
    \xrightarrow{\mathrm{CG}}
    C^{(t)}
    \xrightarrow{\mathrm{Exec./Eval.}}
    \big(\mathbf{x}^{(t)},\mathbf{e}^{(t)}\big), \\
    \big(C^{(t)},\mathbf{e}^{(t)}\big)
    &\xrightarrow{\mathrm{Refine}}
    C^{(t+1)}
    \xrightarrow{\mathrm{Exec./Eval.}}
    \mathbf{x}^{\star}.
\end{aligned}
\label{eq:workflow}
\end{equation}
where $\mathrm{PF}$, $\mathrm{PS}$, $\mathrm{PU}$, and $\mathrm{CG}$ denote the Problem Formulator Agent, Problem Solver Agent, Prompt Upsampler Agent, and Code Generation Agent, respectively. Here, $\mathcal{P}$ represents the structured textual specification of the optimization problem, $\mathcal{A}$ denotes the selected algorithmic strategy, $\Pi$ is the implementation-oriented prompt, $C^{(t)}$ is the generated solver code at the $t$-th refinement step, $\mathbf{x}^{(t)}$ is the corresponding precoding solution, and $\mathbf{e}^{(t)}$ denotes the execution feedback, including runtime status, constraint feasibility, and solution-quality indicators. The final output $\mathbf{x}^{\star}$ is obtained after the code passes execution and evaluation or reaches the refinement termination condition.
\subsection{Domain-Adapted Reasoning Agents}

The first two stages require wireless-specific modeling and optimization knowledge. A general-purpose LLM may parse the task description, but it is not always reliable in identifying communication variables, constraints, convexity properties, or suitable solver structures. Therefore, \emph{AgenticPrecoding} uses two domain-adapted reasoning agents: the \emph{Problem Formulator Agent} and the \emph{Problem Solver Agent}.

The \emph{Problem Formulator Agent} maps $D$ and $\Theta$ into a structured problem specification,
\begin{equation}
    \mathcal{P} = \mathcal{F}_{\mathrm{PF}}(D,\Theta),
\end{equation}
where $\mathcal{F}_{\mathrm{PF}}(\cdot)$ denotes the formulation function. This stage identifies the design variables, optimization objective, feasible domain, and scenario-dependent constraints according to the system configuration, channel information, communication objective, and operational restrictions. Depending on the scenario, the variables may correspond to beamforming matrices, symbol-level transmit vectors, hybrid analog--digital precoders, or auxiliary variables introduced through reformulation. The agent output is a structured textual specification rather than explicit mathematical equations, which provides the necessary modeling information for solver selection.

Given $\mathcal{P}$, the \emph{Problem Solver Agent} selects a suitable non-learning-based optimization strategy,
\begin{equation}
    \mathcal{A} = \mathcal{F}_{\mathrm{PS}}(\mathcal{P}),
\end{equation}
where $\mathcal{F}_{\mathrm{PS}}(\cdot)$ denotes the solver-selection process. This agent examines the objective type, constraint characteristics, convexity, variable coupling, and hardware-related restrictions. According to the identified problem structure, the selected method may involve convex reformulation, semidefinite relaxation, successive convex approximation, alternating optimization, penalty-based optimization, or other classical procedures.

To improve the reliability of these two reasoning stages, we adapt both agents through LoRA-based supervised fine-tuning. The training data are constructed from more than 20 representative references in the precoding literature and converted into task-specific instruction--response pairs for problem formulation and solver selection. This process injects precoding-specific modeling and algorithm-selection knowledge into the agents while preserving the flexibility of LLM-based inference. The solver-selection scope is restricted to non-learning-based methods because they provide clearer mathematical interpretability and are less dependent on fixed training configurations than learning-based precoders.

\subsection{General-Purpose Implementation Agents}

After the problem specification and solver strategy are obtained, the remaining task is to convert them into executable code. Since this stage mainly requires prompt organization, implementation planning, numerical robustness, and programming ability, \emph{AgenticPrecoding} uses two general-purpose implementation agents: the \emph{Prompt Upsampler Agent} and the \emph{Code Generation Agent}.

The \emph{Prompt Upsampler Agent} transforms the scenario descriptor, problem specification, and solver strategy into an implementation-oriented prompt,
\begin{equation}
    \Pi = \mathcal{F}_{\mathrm{PU}}(\Theta,\mathcal{P},\mathcal{A}),
\end{equation}
where $\mathcal{F}_{\mathrm{PU}}(\cdot)$ denotes the prompt upsampling function. The prompt $\Pi$ explicitly organizes the variables, objective, constraints, algorithmic steps, numerical settings, input-output format, and feasibility checks, thereby reducing ambiguity before code generation.

The \emph{Code Generation Agent} then produces executable solver code,
\begin{equation}
    C = \mathcal{F}_{\mathrm{CG}}(\Pi),
\end{equation}
where $\mathcal{F}_{\mathrm{CG}}(\cdot)$ denotes the code-generation process. The generated code implements the selected solver strategy and outputs the corresponding precoding solution after execution. In this way, the implementation agents bridge high-level optimization reasoning and practical solver execution.

\subsection{Feedback-Driven Refinement}

One-shot code generation may still produce syntax errors, runtime failures, infeasible solutions, or poor objective values. To improve robustness, \emph{AgenticPrecoding} introduces a feedback-driven refinement mechanism after code execution.

Let
\begin{equation}
    \mathbf{e}^{(t)} = \mathcal{E}(C^{(t)})
\end{equation}
denote the execution feedback at the $t$-th refinement step, where $\mathcal{E}(\cdot)$ represents the execution and evaluation process. The feedback may include compilation status, runtime status, constraint feasibility, objective value, numerical warnings, and other diagnostic information. Based on this feedback, the code is updated as
\begin{equation}
    C^{(t+1)} = \mathcal{R}\big(C^{(t)},\mathbf{e}^{(t)}\big),
\end{equation}
where $\mathcal{R}(\cdot)$ denotes the refinement operator.

This closed-loop mechanism enables the framework to correct implementation errors, adjust numerical settings, repair constraint violations, and improve solution quality. As a result, \emph{AgenticPrecoding} enhances both the executability of generated code and the feasibility of the obtained precoding solutions across heterogeneous wireless scenarios.
\section{Experiments}
\begin{table*}[!thb]
\centering
\caption{Performance comparison of different methods under nine representative precoding scenarios.}
\label{tab:results}
\scriptsize
\setlength{\tabcolsep}{3pt}
\renewcommand{\arraystretch}{1.12}
\begin{tabular}{c c c l cccccc}
\toprule
\multirow{2}{*}{\textbf{Scenario}} 
& \multirow{2}{*}{\textbf{Optimization Objective}} 
& \multirow{2}{*}{\textbf{Evaluation Metric}} 
& \multirow{2}{*}{\textbf{Method}} 
& \multicolumn{6}{c}{\textbf{SNR (dB)}} \\
\cmidrule(lr){5-10}
& & & & \textbf{0} & \textbf{5} & \textbf{10} & \textbf{15} & \textbf{20} & \textbf{25} \\
\midrule

\multirow{5}{*}{\shortstack{01\\ MU-MIMO}}
& \multirow{5}{*}{\shortstack{Minimize transmit power\\ with rate constraints}}
& \multirow{5}{*}{Power (W) $\downarrow$}
& ZF                    & 0.9140 & 0.1891 & 0.0697 & 0.0275 & 0.0045 & 0.0016 \\
& & & Regularized MMSE      & 2.9181*& 40.0000* & 0.5794 & 0.0662 & 40.0000* & 40.0000* \\
& & & SOCP surrogate      & \underline{0.3253} & \underline{0.0764} & \underline{0.0225} & \underline{0.0072} & \underline{0.0023} & \underline{0.0008} \\
& & & Exhaustive MRT      & 0.3601 & 0.0821 & 0.0240 & 0.0079 & 0.0024 & 0.0009 \\
& & & \textbf{AgenticPrecoding}   & \textbf{0.3144} & \textbf{0.0752} & \textbf{0.0219} & \textbf{0.0065} & \textbf{0.0022} & \textbf{0.0008} \\
\midrule

\multirow{5}{*}{\shortstack{02\\ Constant Envelope \\ precoding}}
& \multirow{5}{*}{\shortstack{Maximize CI margin\\ with unit modulus constraints}}
& \multirow{5}{*}{Normalized Margin $\uparrow$}
& Random CE               & -1.7343 & -3.0840 & -5.4842 & -9.7524 & -17.3425 & -30.8398 \\
& & & MRT + CE projection & 0.3028 & 0.5384 & 0.9574 & 1.7026 & 3.0276 & 5.3839 \\
& & & ZF + CE projection  & 1.5354 & 2.7304 & 4.8553 & 8.6341 & 15.3539 & 27.3036 \\
& & & Coordinate descent CE & \underline{1.7117} & \underline{3.0438} & \underline{5.4127} & \underline{9.6254} & \underline{17.1166} & \underline{30.4381} \\
& & & \textbf{AgenticPrecoding}   & \textbf{1.8076} & \textbf{3.2145} & \textbf{5.7162} & \textbf{10.1651} & \textbf{18.0763} & \textbf{32.1447} \\
\midrule

\multirow{5}{*}{\shortstack{03\\ Hybrid 1-bit\\ MISO}}
& \multirow{5}{*}{\shortstack{Maximize CI margin\\ with 1-bit DAC constraints}}
& \multirow{5}{*}{Normalized Margin $\uparrow$}
& Random 1-bit           & -7.1839 & -12.7752 & -22.7175 & -40.3981 & -71.8391 & -127.7520 \\
& & & ZF 1-bit               & 1.1686 & 2.0781 & 3.6954 & 6.5715 & 11.6859 & 20.7808 \\
& & & CD Greedy 1-bit        & 2.4568 & 4.3689 & 7.7692 & 13.8158 & 24.5684 & 43.6895 \\
& & & MRT 1-bit              & \underline{3.0747} & \underline{5.4677} & \underline{9.7231} & \underline{17.2905} & \underline{30.7473} & \underline{54.6772} \\
& & & \textbf{AgenticPrecoding} & \textbf{3.4530} & \textbf{6.1403} & \textbf{10.9192} & \textbf{19.4174} & \textbf{34.5296} & \textbf{61.4033} \\
\midrule

\multirow{4}{*}{\shortstack{04\\ Cognitive Radio \\ multicast secrecy }}
& \multirow{4}{*}{\shortstack{Maximize minimum \\secrecy rate with \\eavesdropper constraints}}
& \multirow{4}{*}{Secrecy Rate $\uparrow$}
& MRT Beamforming        & 0.1171 & 0.3264 & 0.7587 & 1.3356 & 1.7899 & 2.0176 \\
& & & Secrecy-unaware & 1.0586 & 0.8784 & 2.3706 & 1.4468 & 2.2616 & 3.2564 \\
& & & ZF-like Null-space     & \underline{0.8570} & \underline{1.8339} & \underline{3.1877} & \underline{4.7361} & \underline{6.3596} & \underline{8.0085} \\
& & & \textbf{AgenticPrecoding} & \textbf{1.8265} & \textbf{3.0516} & \textbf{4.3403} & \textbf{5.6406} & \textbf{6.5260} & \textbf{8.6522} \\
\midrule

\multirow{4}{*}{\shortstack{05\\ Full Duplex QAM}}
& \multirow{4}{*}{\shortstack{Minimize transmit power\\ with SINR and SIC constraints}}
& \multirow{4}{*}{Power (W) $\downarrow$}
& MRT                 & 1.0000* & 1.0000* & 0.7500* & 0.7500* & 0.7500* & 0.7500* \\
& & & ZF                  & 0.3420 & 0.2919 & \underline{0.2380} & \underline{0.2072} & \underline{0.2126} & \underline{0.1973} \\
& & & SINR-only           & \underline{0.2637} & \underline{0.2782} & 0.1738* & 0.0502* & 0.1625* & 0.0043* \\
& & & \textbf{AgenticPrecoding}   & \textbf{0.2548} & \textbf{0.2282} & \textbf{0.1987} & \textbf{0.1088} & \textbf{0.0900} & \textbf{0.0673} \\
\midrule

\multirow{6}{*}{\shortstack{06\\ Cognitive Radio \\Sum Rate Maximization}}
& \multirow{6}{*}{\shortstack{Maximize sum rate\\ with transmit power and \\interference limits}}
& \multirow{6}{*}{Sum Rate (bps/Hz) $\uparrow$}
& Random                  & 0.5234 & 0.3511 & 1.1959 & 0.5525 & 1.6311 & 4.5035 \\
& & & Equal power         & 0.2610 & 0.5043 & 1.2395 & 1.4995 & 1.8785 & 1.9926 \\
& & & MRT                 & 2.1662 & 2.3773 & 6.6506 & 6.8911 & 11.4567 & 9.7886 \\
& & & ZF                  & 1.6292 & 1.7958 & 5.6697 & \underline{10.4473} & \underline{28.6721} & \underline{25.9629} \\
& & & Interference-aware & \underline{4.7140} & \underline{5.3194} & \underline{6.6970} & 6.9124 & 12.2773 & 9.3603 \\
& & & \textbf{AgenticPrecoding}   & \textbf{7.1113} & \textbf{9.2167} & \textbf{16.9341} & \textbf{20.0369} & \textbf{36.3210} & \textbf{42.2928} \\
\midrule

\multirow{5}{*}{\shortstack{07\\ Cognitive Radio \\ Inference Robustness}}
& \multirow{5}{*}{\shortstack{Maximize sum rate\\ with interference \\temperature constraints}}
& \multirow{5}{*}{Sum Rate (bps/Hz) $\uparrow$}
& Random      & 4.1639* & 7.6955 & 9.9123 & 15.3307 & 17.1349 & 14.6045 \\
& & & MRT      & \textbf{7.9546} & \underline{10.3148} & \underline{17.0688} & \underline{18.0431} & \textbf{32.9559} & 26.2224 \\
& & & ZF        & 3.3910 & 3.7666 & 4.2445 & 4.0355 & 4.1216 & 3.9976 \\
& & & Interference Nulling & 3.9448 & 6.3498 & 10.1702 & 11.2015 & 23.3777 & \underline{29.3411} \\
& & & \textbf{AgenticPrecoding}    & \underline{4.5831} & \textbf{11.0196} & \textbf{17.6639} & \textbf{23.5504} & \underline{31.9262} & \textbf{33.5494} \\
\midrule

\multirow{6}{*}{\shortstack{08\\ Full Duplex QoS}}
& \multirow{6}{*}{\shortstack{Maximize sum rate\\ subject to total \\power constraint}}
& \multirow{6}{*}{Sum Rate (bps/Hz) $\uparrow$}
& Random                  & 0.5909 & 0.6389 & 1.0991 & 0.7774 & 2.0398 & 1.1365 \\
& & & MRT                 & 4.5349 & 6.8048 & 8.6693 & 10.7856 & 8.2769 & 16.0959 \\
& & & ZF                  & 5.0904 & 8.6230 & 14.6607 & 19.2873 & 22.2449 & 27.7833 \\
& & & RZF                 & \underline{5.6379} & \underline{8.8326} & \underline{14.6892} & \underline{19.2930} & \underline{22.2612} & \underline{27.7841} \\
& & & Plain SLNR          & 5.5980 & 8.8160 & 14.6757 & 19.2907 & 22.2697 & 27.7765 \\
& & & \textbf{AgenticPrecoding}  & \textbf{5.7592} & \textbf{9.1043} & \textbf{14.8177} & \textbf{19.2982} & \textbf{22.4309} & \textbf{27.8034} \\
\midrule

\multirow{5}{*}{\shortstack{09\\ Hybrid Precoding \\ Robustness}}
& \multirow{5}{*}{\shortstack{Minimize transmit power\\ with CI constraints\\ and imperfect CSI}}
& \multirow{5}{*}{Power (W) $\downarrow$}
& Random Hybrid          & 1.0000* & 0.2131 & 0.0719 & 0.0232 & 0.0134 & 0.0039 \\
& & & Phase-Matched Hybrid   & 0.8860 & 0.1581 & 0.0425 & 0.0188 & 0.0067 & 0.0006 \\
& & & Regularized ZF Hybrid  & \underline{0.1981} & 0.0516 & 0.0206 & 0.0065 & 0.0019 & 0.0007 \\
& & & Limited Alternating    & 0.1576 & \underline{0.0440} & \underline{0.0168} & \underline{0.0052} & \underline{0.0017} & \underline{0.0006} \\
& & & \textbf{AgenticPrecoding}      & \textbf{0.0897} & \textbf{0.0210} & \textbf{0.0111} & \textbf{0.0020} & \textbf{0.0007} & \textbf{0.0002} \\
\bottomrule
\end{tabular}

\vspace{1.5mm}
\footnotesize
\textit{Note}: $*$ denotes slightly infeasible results or constraint violations (e.g., $P=40W$ saturation).
\vspace{-2mm}
\end{table*}
\subsection{Experimental Setup}
In this paper, we fine-tune Qwen3-8B on our constructed task-specific datasets to obtain the \textit{Problem Formulator Agent} and the \textit{Problem Solver Agent}. The corresponding training sets contain approximately 6{,}500 and 800 instruction--response pairs, respectively. Both agents are trained using supervised fine-tuning (SFT) with LoRA, where the LoRA rank is set to 8 and the number of training epochs is set to 2. For the remaining two agents, we adopt GPT-5.2 as the \textit{Prompt Upsampler Agent} and Gemini 3.1 Pro as the \textit{Code Generation Agent}.

We evaluate \emph{AgenticPrecoding} on 10 representative precoding scenarios and compare it with several manually selected non-learning-based baselines. These baselines are not chosen by the LLM agents, but are introduced independently to provide external references for evaluating the selected solvers. Learning-based methods are excluded because they are typically tailored to specific scenarios and configurations, making them less suitable for heterogeneous cross-scenario comparison. This setup allows us to assess whether \emph{AgenticPrecoding} can consistently select effective methods and generate stable executable code across diverse precoding problems.

\subsection{Results}
We first evaluate the proposed \emph{AgenticPrecoding} on nine representative precoding scenarios under six SNR levels, namely 0, 5, 10, 15, 20, and 25 dB. The quantitative results are reported in Table~\ref{tab:results}. Overall, \emph{AgenticPrecoding} delivers consistently strong performance across highly heterogeneous problem settings, including power minimization, constructive-interference maximization, secrecy-rate maximization, and sum-rate maximization. In particular, it achieves the best performance in most settings, indicating that the proposed framework can automatically derive effective solvers for substantially different system models, optimization objectives, and constraint structures. As shown in Table~\ref{tab:results}, several conventional baselines become infeasible or numerically unstable in challenging settings, especially under stringent constraints, nonconvex formulations, or hardware-limited designs. By contrast, \emph{AgenticPrecoding} remains markedly more stable across the evaluated scenarios and rarely exhibits the mild constraint violations or severe performance degradation observed in competing methods. Taken together, these results demonstrate that \emph{AgenticPrecoding} is not limited to a narrow class of precoding formulations; instead, it can automatically construct reliable and high-quality optimization procedures with strong cross-scenario adaptability.

\begin{figure}[tb]
	\begin{center}
    \vspace{-2mm} 
		\centerline{\includegraphics[width=\columnwidth]{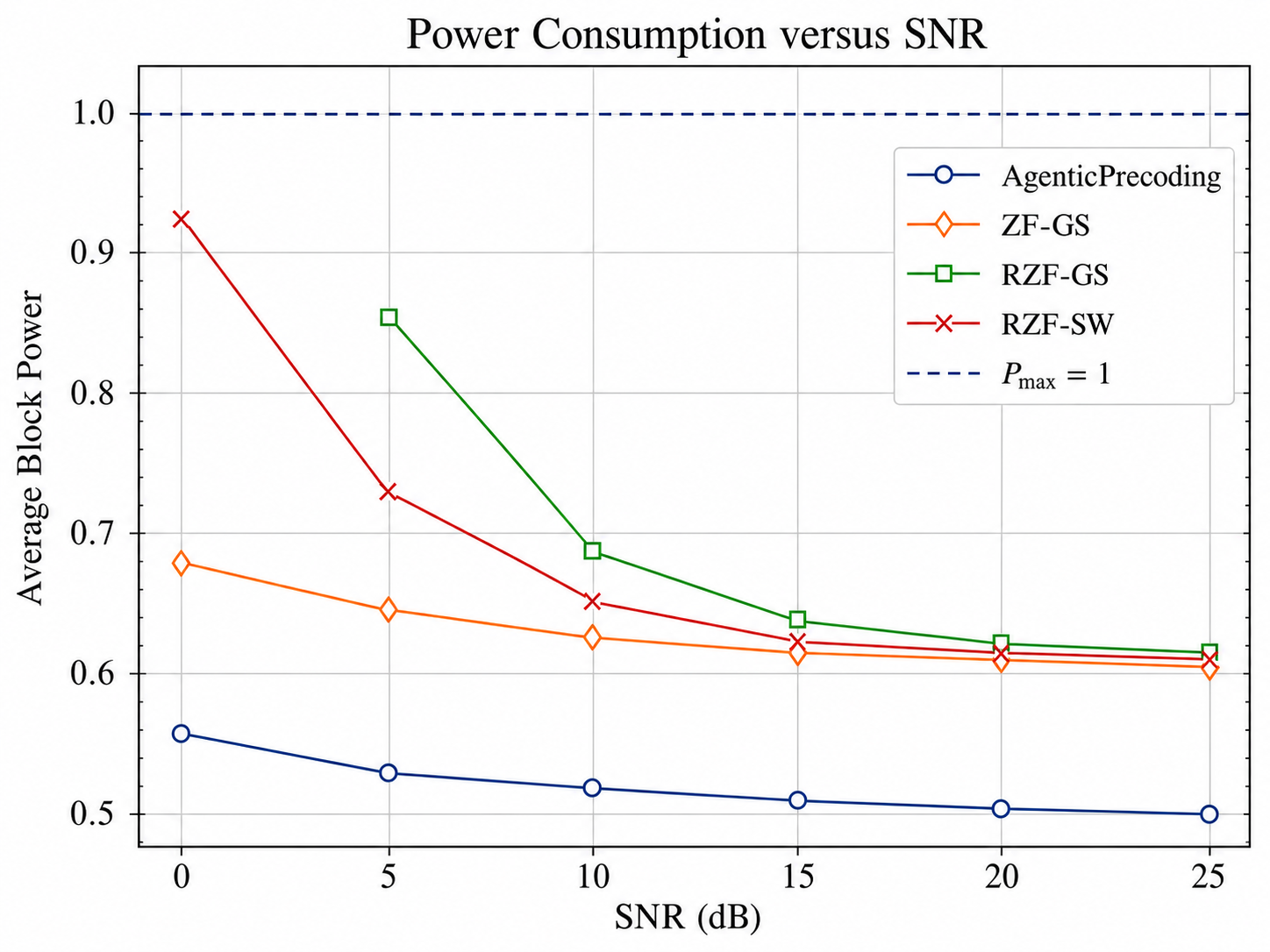}}
		\caption{Average block power consumption versus SNR (dB) for AgenticPrecoding and conventional baselines.}
    \vspace{-2mm} 
		\label{graph3}
	\end{center}
\end{figure}

\begin{figure}[tb]
	\begin{center}
    \vspace{-2mm} 
		\centerline{\includegraphics[width=0.83\columnwidth]{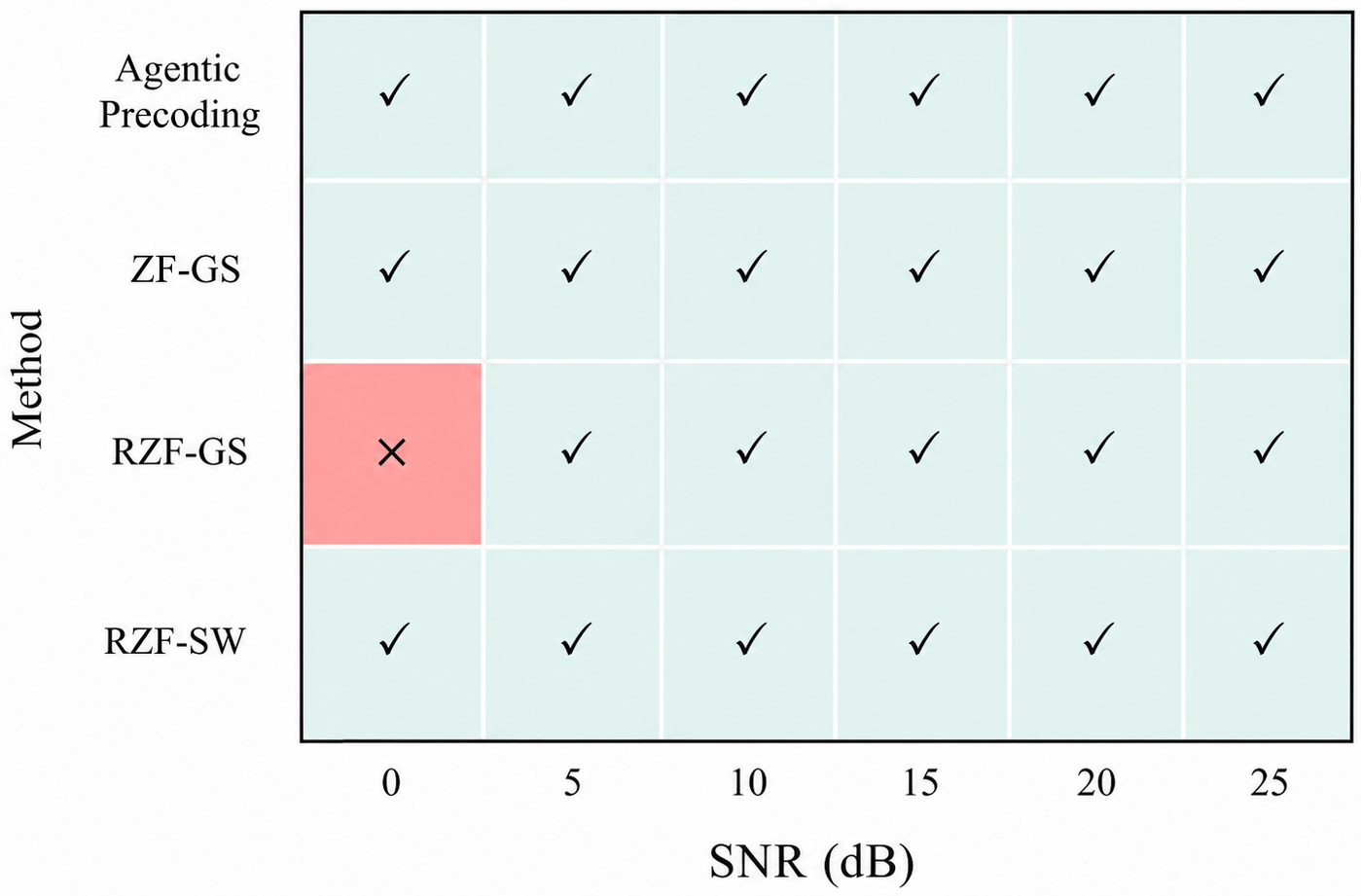}}
		\caption{Feasibility matrix of AgenticPrecoding compared with baseline methods across different SNR levels. }\label{graph4}
	\end{center}
    \vspace{-2mm} 
\end{figure}

To provide deeper insights into the framework's optimization dynamics, we further examine a representative power-minimization case. As shown in \autoref{graph3}, concurring with the broad results in Table~\ref{tab:results}, \emph{AgenticPrecoding} consistently yields the lowest power consumption across all benchmarks. 
Crucially, the implementation reliability of this scenario is validated by the feasibility matrix in \autoref{graph4}. While the RZF-GS baseline fails to converge to a valid solution at 0~dB SNR due to numerical instability, \emph{AgenticPrecoding} maintains a 100\% feasibility rate, demonstrating that \emph{AgenticPrecoding} not only surpasses existing methods in objective optimization but also ensures high operational reliability in complex, real-world wireless environments.

\section{Conclusion}
This paper proposed \emph{AgenticPrecoding}, a universal multi-agent framework that automates end-to-end precoding derivation across heterogeneous wireless environments. By integrating domain-specialized LoRA-tuned agents with a feedback-driven refinement mechanism, the framework overcomes the rigid adaptability and implementation reliability gaps of traditional scenario-specific solvers. Extensive evaluations across diverse precoding scenarios demonstrate that \emph{AgenticPrecoding} consistently achieves superior performance and 100\% feasibility, outperforming conventional baselines particularly in challenging, noise-limited regimes. This study demonstrates the feasibility of multi-agent architectures for physical-layer design, providing an adaptive framework for automated precoding optimization in diverse wireless scenarios.

\bibliography{IEEEabrv,Reference.bib}

@STRING{IEEE_J_SPL        = "{IEEE} Signal Process. Lett."}

@STRING{IEEE_J_COML       = "{IEEE} Commun. Lett."}

@STRING{IEEE_M_COM        = "{IEEE} Commun. Mag."}

@STRING{IEEE_J_TNSE    = "{IEEE} Trans. Netw. Sci. Eng."}

@STRING{IEEE_J_TCOM    = "{IEEE} Trans. Commun."}

@STRING{IEEE_J_TWC     = "{IEEE} Trans. Wirel. Commun."}

@STRING{IEEE_J_IOTJ    = "{IEEE} Internet Things J."}

@STRING{IEEE_J_TCCN    = "{IEEE} Trans. Cogn. Commun. Netw."}

@STRING{IEEE_J_TVT     = "{IEEE} Trans. Veh. Technol."}

@STRING{IEEE_J_TBC     = "{IEEE} Trans. Broadcast."}

@STRING{IEEE_J_STSP    = "{IEEE} J. Sel. Topics Signal Process."}

@STRING{IEEE_J_TSP     = "{IEEE} Trans. Signal Processing"}

@STRING{IEEE_J_SPL     = "{IEEE} Signal Processing Lett."}

@STRING{IEEE_J_COML    = "{IEEE} Commun. Lett."}

@STRING{IEEE_M_COM     = "{IEEE} Commun. Mag."}

@STRING{IEEE_C_GLOBECOM = "{IEEE} Global Commun. Conf. (GLOBECOM)"}

@ARTICLE{DeL4,
  author={Chen, Yijie and Xia, Wenchao and Cai, Shu and Zheng, Gan and Zhu, Hongbo},
  journal=IEEE_J_TNSE, 
  title={Model-Driven Deep Learning-Based Optimization of Downlink Precoding and Fronthaul Compression in Cell-Free {MIMO} Systems}, 
  year={2025},
  volume={12},
  number={3},
  pages={1804-1817},
  doi={10.1109/TNSE.2025.3539824}}

@ARTICLE{DeL3,
  author={Zhu, Wen-Jie and Sun, Chen and Gao, Xiqi and Xia, Xiang-Gen},
  journal=IEEE_J_TWC, 
  title={Deep Learning-Based Precoder Design for Network Massive {MIMO} Transmission}, 
  year={2026},
  volume={25},
  pages={2560-2573},
  doi={10.1109/TWC.2025.3597408}}

@ARTICLE{DeL2,
  author={Zhang, Maojun and Gao, Jiabao and Zhong, Caijun},
  journal=IEEE_J_TWC, 
  title={A Deep Learning-Based Framework for Low Complexity Multiuser {MIMO} Precoding Design}, 
  year={2022},
  volume={21},
  number={12},
  pages={11193-11206},
  doi={10.1109/TWC.2022.3190435}}

@ARTICLE{DeL1,
  author={Hojatian, Hamed and Nadal, Jérémy and Frigon, Jean-François and Leduc-Primeau, François},
  journal=IEEE_J_TWC, 
  title={Unsupervised Deep Learning for Massive {MIMO} Hybrid Beamforming}, 
  year={2021},
  volume={20},
  number={11},
  pages={7086-7099},
  doi={10.1109/TWC.2021.3080672}}

@ARTICLE{llm_precoding_op,
  author={Guo, Wei and Liang, Kai ... [et al.]},
  journal=IEEE_M_COM, 
  title={An {LLM}-Based Framework for Beamforming Optimization}, 
  year={2026},
  volume={64},
  number={4},
  pages={98-104},
  doi={10.1109/MCOM.001.2500487}}

@misc{beamagent,
  title={{BeamAgent}: {LLM}-Aided {MIMO} Beamforming with Decoupled Intent Parsing and Alternating Optimization for Joint Site Selection and Precoding}, 
  author={Wang, Xiucheng and Zhang, Yue and Cheng, Nan},
  year={2026},
  eprint={2603.18855},
  archivePrefix={arXiv}}

@misc{llm-optira,
  title={{LLM-OptiRA}: {LLM}-Driven Optimization of Resource Allocation for Non-Convex Problems in Wireless Communications}, 
  author={Peng, Xinyue and Liu, Yanming and others},
  year={2025},
  eprint={2505.02091},
  archivePrefix={arXiv}}

@INPROCEEDINGS{DL_sumrate,
  author={Chi, Kaiyi and others},
  booktitle=IEEE_C_GLOBECOM, 
  title={{MIMO} Precoding Design with {QoS} and Per-Antenna Power Constraints}, 
  year={2023},
  pages={3324-3329},
  doi={10.1109/GLOBECOM54140.2023.10437214}}

@ARTICLE{DL_low,
  author={Hu, Qiyu and others},
  journal=IEEE_J_TWC, 
  title={Iterative Algorithm Induced Deep-Unfolding Neural Networks: Precoding Design for Multiuser {MIMO} Systems}, 
  year={2021},
  volume={20},
  pages={1394-1410},
  doi={10.1109/TWC.2020.3033334}}

@ARTICLE{DL_power,
  author={Hellings, Christoph and others},
  journal=IEEE_J_TSP, 
  title={Gradient-Based Power Minimization in {MIMO} Broadcast Channels With Linear Precoding}, 
  year={2012},
  volume={60},
  pages={877-890}}

@ARTICLE{DL_reinf,
  author={Tong, Xiao and others},
  journal=IEEE_J_TCOM, 
  title={Symbol-Level Precoding for {MU-MIMO} System With {RIRC} Receiver}, 
  year={2024},
  volume={72},
  pages={2820-2834}}

@ARTICLE{DL_robustness,
  author={Wang, Jiaheng and others},
  journal=IEEE_J_TSP, 
  title={Robust {MIMO} Precoding for Several Classes of Channel Uncertainty}, 
  year={2013},
  volume={61},
  pages={3056-3070}}

@ARTICLE{DL_security,
  author={Kalantari, Ashkan and others},
  journal=IEEE_J_STSP, 
  title={Directional Modulation Via Symbol-Level Precoding: A Way to Enhance Security}, 
  year={2016},
  volume={10},
  pages={1478-1493}}

@ARTICLE{FD_power,
  author={Kabir, Mahmoud T. and others},
  journal=IEEE_J_TWC, 
  title={Interference Exploitation in Full-Duplex Communications: Trading Interference Power for Both Uplink and Downlink Power Savings}, 
  year={2018},
  volume={17},
  pages={8314-8329}}

@ARTICLE{FD_SE_EE,
  author={Teklu, Merhawit Berhane and others},
  journal=IEEE_J_TBC, 
  title={Resource Efficient Full-Duplex Mode of Transmissions Under Imperfect {CSI}}, 
  year={2024},
  volume={70},
  pages={87-98}}

@ARTICLE{CR_inf,
  author={Jung, Minchae and others},
  journal=IEEE_J_COML, 
  title={Interference Minimization Approach to Precoding Scheme in {MIMO}-Based Cognitive Radio Networks}, 
  year={2011},
  volume={15},
  pages={789-791}}

@ARTICLE{CR_robust,
  author={Liu, Lu and others},
  journal=IEEE_J_TWC, 
  title={Robust Symbol Level Precoding for Overlay Cognitive Radio Networks}, 
  year={2024},
  volume={23},
  pages={1403-1415}}

@ARTICLE{CR_security,
  author={Nguyen, Van-Dinh and others},
  journal=IEEE_J_TCCN, 
  title={Enhancing {PHY} Security of Cooperative Cognitive Radio Multicast Communications}, 
  year={2017},
  volume={3},
  pages={599-613}}

@ARTICLE{CR_sumrate,
  author={Nguyen, Van-Dinh and others},
  journal=IEEE_J_TVT, 
  title={An Efficient Precoder Design for Multiuser MIMO Cognitive Radio Networks With Interference Constraints}, 
  year={2017},
  volume={66},
  number={5},
  pages={3991-4004},
  doi={10.1109/TVT.2016.2602844}}

@ARTICLE{CE_CI,
  author={Amadori, Pierluigi Vito and others},
  journal=IEEE_J_TWC, 
  title={Constant Envelope Precoding by Interference Exploitation in Phase Shift Keying-Modulated Multiuser Transmission}, 
  year={2017},
  volume={16},
  pages={538-550}}

@ARTICLE{CE_power,
  author={Liu, Fan and others},
  journal=IEEE_J_SPL, 
  title={An Efficient Manifold Algorithm for Constructive Interference Based Constant Envelope Precoding}, 
  year={2017},
  volume={24},
  pages={1542-1546}}

@ARTICLE{DA,
  author={Wei, Zhongxiang and others},
  journal=IEEE_J_IOTJ, 
  title={Device-Centric Distributed Antenna Transmission: Secure Precoding and Antenna Selection With Interference Exploitation}, 
  year={2020},
  volume={7},
  pages={2293-2308}}

@ARTICLE{ST,
  author={Spano, Danilo and others},
  journal=IEEE_J_TWC, 
  title={Faster-Than-Nyquist Signaling Through Spatio-Temporal Symbol-Level Precoding for the Multiuser {MISO} Downlink Channel}, 
  year={2018},
  volume={17},
  pages={5915-5928}}

@INPROCEEDINGS{HP_CI,
  author={Li, Ang and others},
  booktitle={Proc. EUSIPCO}, 
  title={Hybrid Analog-Digital Precoding for Interference Exploitation}, 
  year={2018},
  pages={812-816}}

@ARTICLE{HP_robust,
  author={Hegde, Ganapati and others},
  journal=IEEE_J_TWC, 
  title={Interference Exploitation-Based Hybrid Precoding With Robustness Against Phase Errors}, 
  year={2019},
  volume={18},
  pages={3683-3696}}
\bibliographystyle{IEEEtran}

\end{document}